\documentclass[prl,aps,twocolumn,showpacs]{revtex4}
\usepackage{graphicx}
\usepackage{bm}

\begin{document}

\title{Scaling relations in anisotropic superconductors with strong pair-breaking}

\author{V. G. Kogan}
\email{kogan@ameslab.gov}
\affiliation{Ames Laboratory and Department of Physics \& Astronomy, Iowa State University, Ames, IA 50011}
\date{Feb.4, 2010} 


\begin{abstract}
Following  the  work of Abrikosov and Gor'kov
 on the pair-breaking effects, one can derive the  temperature
dependences of the electronic specific heat
$C_s/T=\gamma^\prime+\mu T^2$ (with the jump at the superconducting transition $\Delta C \propto T_c^3$) for materials with zero Fermi surface average of the order
parameter  $\langle\Delta\rangle=0$ (e.g. d-wave) or for those with
$\langle\Delta\rangle \ll \Delta_{max}$ (e.g., $\pm s$ of
iron-pnictides) in the presence of strong pair-breaking. Moreover, the
London penetration depth satisfies 
$\lambda^{-2}=\lambda_0^{-2}(1-T^2/T_c^2)$ (or $\lambda -\lambda_0=\beta T^2 $ at low temperatures)  and   the slope of the
upper critical field near $T_c$ is $H_{c2}^\prime \propto T_c$. 
Remarkably simple relations between these at first sight unrelated
quantities take place: $\mu \lambda_0^2
T_c^3/|H_{c2}^\prime|=3\phi_0/8\pi^2$  and $\Delta C\,\beta^2 \,T_c^4/|H_{c2,c}^\prime| =  \phi_0/16\pi^2 $ are  universal constants.   The prediction is checked on two
samples of Ba(Fe$_{1-x}$Co$_{x}$)$_2$As$_2$ and on CeCoIn$_5$ for
which the data needed are available.

\end{abstract}

\maketitle

A remarkable scaling relation between the specific heat jump in
iron-pnictides and their critical temperatures has recently been
established:
$\Delta C\propto T_c^3$ \cite{BNC}. The same relation, in fact,
follows from the theory developed by Abrikosov
and Gor'kov (AG) for isotropic superconductors with magnetic 
impurities  \cite{AG} 
of a concentration close to the critical for which $T_c\ll T_{c0}
$, the critical temperature in the absence of spin-flip
scattering.  This prompted the author to derive the
Ginzburg-Landau (GL) equations for the case of a strongly
anisotropic order parameter with a strong pair-breaking
\cite{K2009}, that made it possible to obtain the jump
\begin{eqnarray}
\Delta C=C_s-C_n=
\frac{16\pi^4k_B^4N(0)\tau_+^2}{3\hbar^2(3\langle\Omega^4\rangle -2)}\, T_c
^3\,.
\label{DeltaC}
\end{eqnarray}
Here, $N(0)$ is the total density of states at the Fermi level
 per one spin. The order parameter  was taken in
the form
$ \Delta ({\bm r},T;{\bm k}_F)=\Psi ({\bm  r},T)\,
\Omega({\bm k_F})$ where  $ \Omega({\bm  k}_F)$   describes the
variation of $\Delta$ along the Fermi surface and is  
normalized to have the average  over the whole  surface  $ \langle
\Omega^2  \rangle=1$. 

Another assumption  employed in the
derivation of Eq.\,(\ref{DeltaC}) was that 
$\langle\Delta\rangle=\langle\Omega\rangle=0$ (which is true for
the d-wave) or  that
$\langle\Delta\rangle\ll \Delta_{max}$ (presumably true for
iron-pnictides with $\pm s$ order parameter). 
 
The  scattering in the Born approximation was characterized
by two scattering times, the transport $\tau $   responsible for
the normal conductivity and  $\tau_m$ for processes breaking the
time reversal symmetry (e.g., spin-flip):
\begin{equation}
 1/\tau_+ = 1/\tau + 1/\tau_m  \,.
  \label{taus}
  \end{equation}
 This is of course a gross simplification. For multi-band
Fermi surfaces one may need more parameters for various intra- and
inter-band processes; these, however, are hardly controllable and
their number is too large for  a useful theory.   

It is well-known that the formal  AG scheme for uncorrelated 
magnetic impurities \cite{AG} applies to various situations with
different pair-breaking causes, not necessarily the AG spin-flip
scattering \cite{Maki}.  A  particular  example relevant for
iron-pnictides is antiferromagnetic superconductors in which even
non-magnetic impurities cause pair-breaking, see e.g. review
\cite{BB}. 

Another quantity obtained in \cite{K2009} from the GL equations is the slope of $H_{c2}$   at $T_c$:
      \begin{equation}
\frac{dH_{c2,c}}{dT}  = -\frac{4\pi \phi_0 k_B^2 }{ 3\hbar^2\langle
\Omega^2 v_a^2    \rangle}\,T_c  \,  
\label{slope}
\end{equation}
for the field along the $c$ axis of a uniaxial crystal   ($\phi_0$ is the flux quantum).  The scaling relation $H_{c2}^\prime \propto T_c$ holds across the whole family of iron-pnictides, albeit with a considerable scatter; a number of data points are collected in \cite{K2009} to support this statement.   
There are reasons for this scatter:  different
criteria in extracting $H_{c2}$ from resistivity data,  
unavoidable uncertainties rooted in sample inhomogeneities in
determination of $T_c$ and the slopes of $H_{c2}(T)$ near $T_c$,     possible
differences in Fermi velocities and the order parameter anisotropies, to name a few. Since  the model employing 
only two scattering parameters for  multi-band iron-pnictides is a far-reaching simplification,   one can expect the model to work  qualitatively at best.  

The third piece of evidence in favor of the strong pair-breaking in iron-pnictides is
 the low temperature behavior of the London penetration depth 
$\lambda(T)$.   Evaluation of $\lambda(T;\tau,\tau_m)$ for arbitrary $\tau$'s
and arbitrary anisotropy of $\Delta$ is difficult analytically.
However, for a strong $T_c$ suppression, 
as was shown by AG \cite{AG},  the  formalism used for  derivation of the
GL equations near $T_c$ applies at {\it all temperatures}. Physically, this is because the pair-breaking suppresses the order parameter so that the expansion in powers of $\Delta$ and its derivatives can be done at any $T$.  The calculation then
proceeds in a  manner similar to that near $T_c$ \cite{lambda-prl}. Within the same formal scheme that gave  Eq.\,(\ref{DeltaC}) for the specific heat jump and Eq.\,(\ref{slope}) for the $H_{c2}$ slopes, one obtains for the order parameter  in the field-free state:
\begin{eqnarray}
 \Psi^2 =\frac{2\pi^2k_B^2(T_c^2 -T^2)}
{3\langle\Omega^4\rangle-2  } \,;
\label{Psi}
\end{eqnarray}
  this reduces to the AG form for $\Omega=1$. The free energy difference 
  between the normal and superconducting states is given by \cite{K2009}:
\begin{eqnarray}
F_n-F_s=    \frac{2\pi^4N(0)\tau_+^2}{3\hbar^2(3\langle\Omega^4\rangle
-2)}\,k_B^4 (T_c^2 -T^2)^2\, .
\label{F5}
\end{eqnarray}
Hence, one can evaluate not only the jump $\Delta C$ at $T_c$, but
the full temperature dependence
$C_s(T)=-T\partial^2 F_s/\partial T^2$:
\begin{eqnarray}
\frac{C_s}{T}&=& (\gamma-\mu T_c^2/3) +\mu \,T^2\,,\label{C}\\
\mu&=& \frac{8\pi^4k_B^4N(0)\tau_+^2}{\hbar^2(3\langle\Omega^4\rangle-2)} \, .
\label{mu}
\end{eqnarray}
Here, $\gamma T$ is the electronic   heat capacity  of the normal phase. The dependence of the type (\ref{C}) has recently been reported in Refs.\,\cite{meingast,sefat,Wen}  for Ba(Fe$_{1-x}$Co$_x$)$_2$As$_2$.

Within the same model one can also calculate  the London penetration depth \cite{lambda-prl}:
\begin{equation}
 (\lambda^2)_{ik}^{-1}= \frac{16\pi^3 e^2N(0)k_B^2\tau_+^2 }{ c^2 \hbar^2(3\langle\Omega^4\rangle-2)}
\Big\langle v_iv_k \Omega
 ^2\Big\rangle(T_c^2-T^2)\,.
    \label{lambda_gapless}
\end{equation}
The low $T$ behavior of $\Delta\lambda_{ab}=\lambda_{ab}(T)-\lambda_{ab}(0) $ for a
uniaxial material is given by:
\begin{eqnarray}
\Delta\lambda_{ab}=\frac{\eta}{T^3_c}T^2,\, \,\,\,
\eta = \frac{c\hbar}{8\pi k_B \tau_+}\sqrt{\frac{
 3\langle\Omega^4\rangle-2 }{ \pi
e^2N(0)\langle v_a^2 \Omega
 ^2 \rangle } }.
    \label{dlambda}
\end{eqnarray}
The power-law behavior of $\Delta\lambda_{ab}\propto T^n$ with $n\approx 2$ has been seen in many iron-pnictides \cite{lambda-prl}. Moreover, the pre-factor $\eta/T_c^3$ extracted from the data varies from one material to another approximately as $1/T_c^3$  that further supports the pair-breaking scenario. 
One readily obtains for $T=0$:
\begin{eqnarray}
 \lambda_{ab}(0)= 2\eta/T _c \,.
     \label{lambda(0)}
\end{eqnarray}

The relations (\ref{C}),(\ref{mu}) for the specific heat,
(\ref{slope}) for the slope of $H_{c2}$, and (\ref{lambda_gapless})--(\ref{lambda(0)})
for the penetration depth contain material parameters: the density
of states $N(0)$, Fermi velocities $v_i$, the parameter $\Omega$ of the gap
anisotropy, and the combined scattering time $\tau_+$. One can 
express the ratio $(3\langle\Omega^4\rangle-2)/\tau_+^2$ in
terms of the coefficient $\mu$ of
Eq.\,(\ref{C}) and substitute the result in the expression for
$\lambda(0)$, thus establishing a relation between 
 the behavior of the specific heat and the penetration depth. $\langle v_a^2 \Omega
 ^2 \rangle $ can in turn be expressed in terms of the $H_{c2}$ slope of Eq.\,(\ref{slope}). Surprisingly, this simple algebra yields
\begin{eqnarray}
\frac{\mu \,\lambda_{ab}^2(0)\,T_c^3}{|H_{c2,c}^\prime|}= \frac{3\phi_0}{8\pi^2} 
     \label{universe}
\end{eqnarray}
with a universal quantity on the right-hand side (RHS). Four 
quantities on the LHS are measured in independent experiments so
that this result might be used as a stringent test of the
pair-breaking scenario.  It is worth reiterating that this relation
is obtained for 
  nearly zero $\langle\Delta\rangle$'s in materials with  a strong
pair-breaking. 

To check the scaling relation (\ref{universe}) one can turn to data on 
Ba(Fe$_{1-x}$Co$_{x}$)$_2$As$_2$ with $x=0.075$ and $T_c=21.4\,$K for which the behavior of
the electronic specific heat has recently been  reported
\cite{meingast}.  The phonon background was evaluated by measuring the specific heat of an overdoped material with $x=0.153$ for which superconductivity is suppressed.  
This procedure yields $C_s(T)$ of the form (\ref{C}) with 
\begin{eqnarray}
 \mu\approx 0.12\, {\rm mJ/mol\,K} ^4
\approx
20\,{\rm erg/cm} ^3 {\rm K} ^4 \,.
\label{zeta}
\end{eqnarray}
There is still a considerable uncertainty in experimental 
values for  zero-$T$ penetration depth  which in various studies with different techniques is 
found to be   $\lambda_{ab}(0)\approx (270\pm 50)\,$nm \cite{Williams,KAM}.
Taking 
$H_{c2,c}^\prime\approx 2.5\,$T/K, see \cite{K2009}, one estimates:
\begin{eqnarray}
\frac{\mu
\,\lambda_{ab}^2(0)\,T_c^3}{|H_{c2,c}^\prime|}\approx (3.7 - 7.8)\times
10^{-9}{\rm G\,cm}^2\,.
\label{estimate1}
\end{eqnarray}
This is of the same order of magnitude as
the RHS of Eq.\,(\ref{universe}): $3 \phi_0/8\pi^2 \approx
7.6\times 10^{-9}{\rm G\,cm}^2$.

Similar specific heat results have recently been reported for the same compound with $x=0.08$, albeit with a different way to subtract the lattice contribution \cite{sefat}. The low temperature behavior of $C_s/T$ is linear in $T^2$ with approximately the same value of the slope $d(C_s/T)/d(T^2)$ as reported in \cite{meingast}.

 Since other iron-pnictides comply with the scaling relations  (\ref{DeltaC}),  (\ref{dlambda}), and (\ref{slope}) (see \cite{BNC}, \cite{lambda-prl} and \cite{K2009}), they are good candidates to obey the relation (\ref{universe}).  Unfortunately, for a precious few of these compounds measurements of all quantities involved in Eq.\,(\ref{universe}) were done on the same samples. A good example is Ref.\,\cite{Wen} where $C_s/T$ of a few crystals of Ba(Fe$_{1-x}$Co$_{x}$)$_2$As$_2$   was found to be linear in $T^2$ ($C(T)$ in the field of 9\,T was taken as the phonon contribution and was subtracted). However, neither  $H_{c2}^\prime(T_c)$ nor $\lambda(0)$ are available for these samples. In this situation one can use a   crude estimate of  slopes $H_{c2}^\prime(T_c)$ based on comparison of the scaling relation (\ref{slope}) with data on many iron-pnictides \cite{K2009}, $H_{c2}^\prime(T_c)\approx (0.2\,{\rm T/K}^2)\,T_c$ to obtain values listed in Table 1. Then, with the help of Eq.\,(\ref{universe}) one can calculate the penetration depth. The values of $\lambda(0)$ so obtained and given in the Table are quite reasonable.
 
  \begin{table}[htb]
\caption{The penetration depth $\lambda(0)$ calculated with the help of Eq.\,(\ref{universe}) for Ba(Fe$_{1-x}$Co$_{x}$)$_2$As$_2$. The slopes $\mu=d(C_s/T)/d(T^2)$ are extracted from Fig.\,3b of Ref\,\cite{Wen}. $T_c$ is taken as corresponding to zero resistance. $H_{c2}^\prime(T_c)$ is evaluated as explained in the text.}
 \label{table1}
\begin{ruledtabular}
\begin{tabular} {llccc}
x & $T_c\,$K & $\mu\,$mJ\,/K$^4$  & $H_{c2}^\prime(T_c)\,$T/K &  $\lambda(0)\,$nm\\
\hline

0.06 & 14.5   & 0.8    & 2.9  &  240 \\
0.07 & 19.2   & 0.67   & 3.8   & 200  \\
0.08 & 23.9   & 0.73   & 4.8  & 150 \\
0.12 & 17.3   & 0.75   & 3.4     &200 \\
0.15 & 7.0  & 0.77     & 1.4     & 500\\
\end{tabular}
\end{ruledtabular}
\end{table}

 As mentioned, the results discussed above hold for strong pair-breaking in materials with the order parameter  averaged over the Fermi surface close to zero. This is exactly the case for the d-wave compounds. Hence, the underdoped cuprates with a strongly suppressed $T_c$ may also be among materials satisfying the above scaling formulas and the scaling relation (\ref{universe}), in particular. 
It is worth noting that the scaling (\ref{lambda(0)}) is supported by 
 the surface resistance \cite{Bonn} and  optical data
\cite{Homes}  for  a series of samplesof  YBa$_2$Cu$_3$O$_{6+x}$ 
with $T_c$ varying from 3 to 17\,K.  

Another example where the model developed here may work is CeCoIn$_5$. 
 This is a clean  heavy-fermion  d-wave superconductor \cite{petrovic,dwave}  with $T =2.3\,$K; 
in the normal phase, the
material is a  paramagnet \cite{Ames}. Note that  quantities entering the relation (\ref{universe}) are either for zero-field ($\mu$ and $\lambda(0)$) or for temperatures near $T_c$ ($H_{c2,c}^\prime$), so that the domain of interest here is not affected by complications related to 
  paramagnetic constrains or to possible Fulde-Ferrell-Larkin-Ovchinnikov  phase. 
  
  The temperature dependence of the penetration depth of CeCoIn$_5$
\cite{115} is well described by
$\lambda=\lambda(0)/\sqrt{1-T^2/T_c^2}$ (with $\lambda_0\approx
360\,$nm) \cite{Ames} as is should be for a strong pair-breaking,  
Eq.\,(\ref{lambda_gapless}). Moreover,  $C_s/T$ (where $C_s$ is the
electronic part of the specific heat) is close to being linear in
$T^2$ at low temperatures \cite{115_C(T)} in agreement with
Eq.\,(\ref{C}). Reading the data on $C_s/T$ on Fig.\,2b of
\cite{115_C(T)}, one can estimate the coefficient $\mu$ of the
$T^2$ part as $\mu\approx 3.5\times
10^4\,$erg/cm$^3$K$^3$. The slope of the upper critical field is
known too: $H_{c2,c}^\prime\approx 11.5\,$T/K. Then, the LHS of the
scaling relation (\ref{universe}) is estimated as $4.8\times
10^{-9}\,$G\,cm$^2$, reasonably close to the universal number 
$7.6\times 10^{-9}\,$G\,cm$^2$ on the RHS of this relation.  

The scaling relation (\ref{universe}) involves two quantities,
the slope $\mu=d(C_s/T)/d(T^2)$  and the zero-$T$ penetration depth $\lambda(0)$,
both of which are difficult to access experimentally (the first
because of the necessity to subtract the phonon contribution from
the measured $C/T$,  the second because usually only the
deviation of $\lambda(T)$ from $\lambda(0)$ is measured).
Fortunately, one can avoid these difficulties using another scaling
relation. Writing Eq.\,(\ref{dlambda}) as $\Delta\lambda_{ab}=\beta
T^2$
 with $\beta=\eta/T_c^3$ one readily finds:
\begin{eqnarray}
\frac{\Delta C\,\beta^2 \,T_c^4}{|H_{c2,c}^\prime|}=
\frac{ \phi_0}{16\pi^2} = 1.27\times 10^{-9}\,{\rm G\,cm^2}.
     \label{universe1}
\end{eqnarray}
Measurement of the jump $\Delta C$ does not require the phonon contribution be subtracted and determination of $\beta = d(\Delta\lambda)/d(T^2)$ does not require knowledge of $\lambda(0)$.

It is worth noting a couple of other scaling relations which might be useful. From expressions (\ref{DeltaC}) and (\ref{mu}) immediately follows:
\begin{eqnarray}
 \Delta C =
\frac{ 2T_c^2}{3}\,\mu \,.
     \label{DC-mu}
\end{eqnarray}
Similarly, Eqs.\,(\ref{dlambda}) and (\ref{lambda(0)}) yield:
\begin{eqnarray}
 \lambda(0) =
\frac{ 2T_c^2}{T^2}\,\Delta\lambda(T) \,.
     \label{lam0-Dlam}
\end{eqnarray}

A word of caution here is in order. The scaling relations discussed are obtained for superconductors with 
 $\langle\Delta\rangle\approx 0$ and in the limit of  strong pair-breaking, strictly speaking for $T_c\to 0$. It is shown that in this limit $C_s/T$ is linear in $T^2$ and $\Delta\lambda$ is proportional to $T^2$. For  systems discussed above, this is only approximately so. E.g., for CeCoIn$_5$ 
  the actual fit of the
experimental $C_s/T$ \cite{115_C(T)} to the form   $\mu_1+\mu T^\nu$ yields
$\nu\approx 2.3$  indicating that the material is not quite in
the limit of a strong pair-breaking.  Similarly, the data \cite{lambda-prl} yield the power-law for $\Delta\lambda_{ab}\propto T^n$ with $n$ close but not exactly equal to 2, so that the iron-pnictides examined are also not exactly in the strong pair-breaking limit. This, along with common experimental uncertainties, is, probably, the main source of deviations from the scaling relations proposed.
 
To summarize, scaling relations are found involving the specific heat, the penetration depth, and the upper critical field for superconductors with a strong pair-breaking and the averaged over the Fermi surface order parameter close to zero (as is the case for d-wave or for $\pm s$ symmetries with impurities). The relation are checked on a few iron-pnictides and on CeCoIn$_5$. 
 
I am grateful to my coleagues S.~Bud'ko, R.~Prozorov, P.~Canfield,  C.~Petrovic, J.~Clem, and J.~Schmalian  for interest and help. Work at the Ames Laboratory was supported by the
Department of Energy - Basic Energy Sciences under Contract No. DE-AC02-07CH11358.

 \end{document}